\documentclass[12pt]{article}
\pdfoutput=1
\usepackage{amssymb,amsmath,slashed,amscd}
\usepackage[pdftex]{graphicx}
\makeatletter

\@addtoreset{equation}{section}
\def\section{\@startsection {section}{1}{\z@}{-2.5ex plus -1ex minus
 -.2ex}{1.3ex plus .2ex}{\large\bf}}
\def\subsection{\@startsection{subsection}{2}{\z@}{-2.25ex plus%
 -1ex minus -.2ex}{0.5ex plus .2ex}{\bf}}

\advance \voffset by -0.8in
\advance \hoffset by -0.9in
\newcommand{\CC}{\mathbb{C}}
\newcommand{\RR}{\mathbb{R}}
\newcommand{\ZZ}{\mathbb{Z}}
\newcommand{\NN}{\mathbb{N}}
\newcommand{\bee}{\begin{equation}}
\newcommand{\eee}{\end{equation}}
\def\Dslash{\slashed{D}}

\begin{document}

\begin{flushright}
EMPG-16-10
\end{flushright}
\vskip 10pt
\baselineskip 28pt

\begin{center}
{\Large \bf  Spectral Properties of  Schwarzschild  Instantons}

\baselineskip 18pt

\vspace{0.4 cm}

{\bf Rogelio Jante and Bernd J.~Schroers}\\
\vspace{0.2 cm}
Maxwell Institute for Mathematical Sciences and
Department of Mathematics,
\\Heriot-Watt University,
Edinburgh EH14 4AS, UK. \\
{\tt rj89@hw.ac.uk} and {\tt b.j.schroers@hw.ac.uk} \\

\vspace{0.2cm}

{ April  2016} 
\end{center}

\begin{abstract}
\noindent We study spectral properties of the Dirac and scalar  Laplace operator on the Euclidean Schwarzschild space, both twisted by a family of  abelian connections with anti-self-dual curvature.  We show that the zero-modes of the gauged  Dirac operator,  first studied  by  Pope,  take a particularly simple form in terms of the radius of the Euclidean time orbits, and interpret them in the context of geometric models of matter.  For the gauged Laplace  operator, we study the spectrum of  bound states numerically and observe that it can be approximated with remarkable accuracy by that of the exactly solvable gauged Laplace  operator on the Euclidean Taub-NUT space. 
\end{abstract}

\baselineskip 16pt
\parskip 3 pt
\parindent 0pt

%
%

\section{Introduction}

The Euclidean Schwarzschild (ES)   geometry  is one of the simplest and best-studied solutions of the Euclidean Einstein equations in four dimensions. It has also been known for a long time that it supports a family of  abelian instantons, i.e.,  (anti-) self-dual solutions of the (Euclidean) Maxwell equations. The purpose of this paper is to explore the effect of minimally coupling this family to the Dirac and scalar Laplace operator  on the spectrum of these operators.

The ES space was  first considered  by Hawking  in \cite{Hawking} as an example of a gravitational instanton. The transition from the usual Lorentzian to the Euclidean Schwarzschild geometry involves a Wick rotation into a imaginary time direction and, crucially, the compactification  of the Euclidean time to a circle.

Topologically, the ES space is a product of the  two-sphere  $S^2$ and the two-plane, but in anticipation of the metric structure we think of the latter as the  open disk $D^2$. The isometry group of the ES metric is  $O(3)\times O(2)$, with $O(3)$ acting  in the  standard fashion on  $S^2$, and $O(2)$ acting naturally on the disk, fixing  its  origin $O$.  In the terminology of \cite{GH}, the two-sphere  $\{O\}\times S^2$ of fixed points is a bolt; away from the  bolt,  ES space  has  the structure of a trivial circle bundle. 

In the context of the recently proposed geometric models of matter \cite{AMS},   non-compact four-manifolds which are asymptotic circle  fibrations are interpreted as models for particles, with  the  Chern number of the asymptotic bundle modelling  a particle's electric charge. In \cite{AMS}, only  compact four-manifolds were considered for neutral particles,   but once non-compact manifolds are included, the ES space emerges as a natural candidate model for the neutron \cite{AFS}. While this remains a  speculative proposal, it triggered the work presented here.

The main motivation of this paper, however,  lies in the intrinsic interest of the spectral properties of gravitational instantons, particularly those which are asymptotically locally flat (ALF), like the Taub-NUT (TN) geometry, the Atiyah-Hitchin manifold or the ES space.  All of these support abelian instantons, and the spectral properties of the Dirac and scalar Laplace operators on these spaces, minimally coupled to the abelian gauge field, are very rich. The scalar Laplace operators display all the phenomena known from standard three-dimensional quantum mechanics - bound states, resonances and scattering - but do so in a completely geometrical and smooth context, see \cite{GM, Manton, Schroers, JS,JS2} for details. 

Among the gravitational instantons, the TN space plays a role akin to that of the hydrogen atom in atomic physics. Topologically, it is simply $\RR^4$, but, away from the origin, it is useful to think of it as a non-trivial disk bundle over $S^2$, with the fibre collapsing at the origin (the nut).  There is a dynamical symmetry which allows  for the  spectrum to be determined exactly. Moreover, coupling the abelian instanton leads to  a spectral problem which combines the interesting  features of  Landau levels `in the fibre'  with a  `Coulomb problem in the base' \cite{JS2}. 

An important theme of this paper is the similarity between the spectral properties of the TN  space and those of the ES space, both coupled to their respective abelian instantons. The similarities can all be traced back to a simple geometrical fact, which does not appear to have been pointed out in the literature: the metrics  of  the disk fibres in  TN  (where they are twisted) and  in  ES (untwisted)  are the {\em  same}. Qualitatively, the fibre metric is that of a cigar-shaped submanifold of Euclidean three-space.

To end this introduction, we summarise the organisation of the paper together with our main results.  In Sect.~2, we review the derivation of the ES geometry. We give its form when the proper radial distance on the disk is chosen as a radial coordinate, and point out the isometry between the fibres in the TN and ES spaces. We also review the family of abelian instantons on the ES space, first discussed in \cite{Pope1} and recently in more detail in \cite{Etesi}. These instantons are  labelled by an integer and play a key role in our discussion. Restricted to any of the  two-spheres, they are essentially Dirac monopoles, and the integer is the magnetic charge, denoted $p$ in this paper.  

In Sect.~3, we revisit the result due to Pope \cite{Pope1} that the Dirac operator twisted by an abelian instanton of charge $p$ has a kernel of dimension $|p|^2$.  We exhibit the zero-modes in complex coordinates on $S^2$ and the proper radial distance on $D^2$, and show that they take a particularly simple, essentially holomorphic form. 
In geometric models of matter, zero-modes of Dirac operators on four-manifolds are potential candidates for describing the  spin degrees of freedom of an elementary particle. We  therefore decompose the kernel of the Dirac operator into irreducible representations of the  spin group $SU(2)$, and show that, for magnetic charge $p$, the kernel contains precisely $|p|$ copies of the $|p|$-dimensional representation. 

Sect.~4 contains our results on the spectrum of the gauged scalar Laplace operator. We show that, unlike  in the  ungauged case studied in \cite{Fawcett}, the gauged Laplace operator has bound states, and compute their eigenvalues numerically. We show that the radial eigenvalue problem can be mapped onto the radial eigenvalue problem in  the TN case, and, in this way, obtain an approximate formula for the eigenvalues which agrees with the numerically calculated eigenvalues to a remarkable accuracy (less than $0.01 \%$ discrepancy for the 11th or higher eigenvalues).

The final Sect.~5 contains a discussion of our results, while the Appendix gives a description of monopole spherical harmonics as local  sections of suitable powers of the hyperplane bundle over $S^2$, which we use throughout the paper.

\section{Euclidean Schwarzschild Geometry}

\subsection{Vacuum Einstein equations}\label{one}

Natural coordinates  on the ES space which make  its $O(3)\times O(2)$ symmetry manifest are spherical coordinates $(\theta, \phi)$  on $S^2$ and polar coordinates $\tau$ (the periodic Euclidean time) and $r$ on $D^2$. In terms of these, the metric takes the form
\begin{equation}
\label{schwarMetric1}
ds^2 = a^2(d\theta^2 + \sin^2\theta d\phi^2) + f^2dr^2 + c^2d\tau^2,
\end{equation}  
where $a, c$ and $f$ are functions of $r$. The function $f$  can be chosen to fix the  radial coordinate  $r$, but  $a$ and $c$  are  then  determined as functions of $r$  by the vanishing of the Ricci tensor. Geometrically, $a$  determines the size of the $O(3)$ orbits and $c$  the size of the $O(2)$ orbits. The function $c$ vanishes on the bolt,   where $a$ takes its smallest value.  The bolt is the Euclidean analogue of the horizon of the Schwarzschild black hole \cite{Wald}. 

For our study of the Dirac and Laplace operator on the ES  space we require a detailed understanding of the Riemannian geometry and  the spin connection. We therefore briefly go through the solution of the Einstein equations, emphasising  the choice of $c$ as radial coordinate. Even though this is geometrically natural, it does not appear to have been considered in the literature. It is useful for us because we are interested in  the comparison with the TN geometry, which was studied in analogous coordinates in \cite{AFS}.

We begin without fixing a radial  or Euclidean time  coordinate in \eqref{schwarMetric1} and  choose  the vierbein
\begin{equation}
e^1 = ad\theta, \ \ e^2 = a\sin\theta d\phi, \ \ e^3 = fdr, \ \ e^4 = cd\tau,
\end{equation}
so  that the dual vector fields,  defined via $e^{\alpha}(E_{\beta}) = \delta^{\alpha}_{\beta},
\;  \alpha,\beta =1,\ldots, 4$,    are 
\begin{equation}
E_1 = \frac{1}{a}\partial_{\theta}, \ \ E_2 = \frac{1}{a\sin\theta}\partial_{\phi}, \ \ E_3 = \frac{1}{f}\partial_r, \ \ E_4 = \frac{1}{c}\partial_{\tau}.
\end{equation} 
Calculating the  components of the spin connection one-forms  $ \omega^{\alpha}_{\ \beta}$   from  $de^{\alpha} + \omega^{\alpha}_{\ \beta}\wedge e^{\beta}=0$ (where the indices are placed up or down for convenience)  and  $ \omega^{\alpha}_{\ \beta} = -\omega^{\beta}_{\ \alpha}$, we find the non-vanishing components
\begin{equation}
\label{ESconnection}
\omega^3_{\ 1} = - \frac{a'}{af}e^1, \ \ \omega^3_{\ 2} = -\frac{a'}{af}e^2, \ \ \omega^1_{\ 2} = -\frac{\cos\theta}{a\sin\theta}e^2, \ \ \omega^4_{\ 3} = \frac{c'}{cf}e^4,
\end{equation}
so that the non-vanishing Riemann  curvature two-forms $R^{\alpha}_{\ \beta} = d\omega^{\alpha}_{\ \beta} + \omega^{\alpha}_{\ \gamma}\wedge\omega^{\gamma}_{\ \beta}$ come out as 
\begin{align}
\label{EScurv}
R^{1}_{\ 2} &= \frac{1}{a^2}\left[1 - \left(\frac{a'}{f}\right)^2\right]e^1\wedge e^2, \quad R^{3}_{\ 1} = -\frac{1}{af}\frac{d}{dr}\left(\frac{a'}{f}\right)e^3\wedge e^1, \\
R^{3}_{\ 2} &= -\frac{1}{af}\frac{d}{dr}\left(\frac{a'}{f}\right)e^3\wedge e^2,  \quad  R^{4}_{\ 1} = -\frac{a'c'}{acf^2}e^4\wedge e^1, \nonumber \\
R^{4}_{\ 2}& = -\frac{a'c'}{acf^2}e^4\wedge e^2, \quad R^{4}_{\ 3} = -\frac{1}{cf}\frac{d}{dr}\left(\frac{c'}{f}\right)e^4\wedge e^3.  \nonumber 
\end{align}
It follows that the  Ricci tensor $R_{\alpha\beta} = R^{\gamma}_{\ \alpha\gamma\beta}$  is  diagonal.
 Using  spherical symmetry (or explicitly from \eqref{EScurv})  we note that   $R_{11} = R_{22}$, and so that the vanishing of the Ricci tensor gives  three independent equations, from the vanishing of $R_{11}, R_{33} $ and $R_{44}$:
\begin{align}
\label{Einsteineq}
\frac{1}{af}\frac{d}{dr}\left(\frac{a'}{f}\right) + \frac{a'c'}{acf^2} - \frac{1}{a^2}\left(1 - \left(\frac{a'}{f}\right)^2\right) &= 0,\\
\frac{1}{cf}\frac{d}{dr}\left(\frac{c'}{f}\right) + \frac{2}{af}\frac{d}{dr}\left(\frac{a'}{f}\right) &= 0, \nonumber \\
\frac{1}{cf}\frac{d}{dr}\left(\frac{c'}{f}\right) + \frac{2a'c'}{acf^2} &= 0. \nonumber 
\end{align}
From the last two equations we deduce 
\begin{equation}
\label{schwar0}
\frac{d}{dr}\left(\frac{a'}{f}\right) = \frac{a'c'}{cf},
\end{equation}
which can be integrated to give 
\begin{equation}
\label{schwar1}
\frac{a'}{f} = \frac{c}{\Lambda},
\end{equation}
where $\Lambda$ a constant of dimension length. Inserting this into the first equation in \eqref{Einsteineq} yields
\begin{equation}
\label{schwar2}
\frac{2a}{f}\frac{c'}{\Lambda} = 1-\frac{c^2}{\Lambda^2}.
\end{equation}

We have obtained  the two first order  equations \eqref{schwar1} and \eqref{schwar2}  from the three second order equations \eqref{Einsteineq}, but in fact the two equations \eqref{schwar1} and \eqref{schwar2}  are equivalent to \eqref{Einsteineq}. To show this, solve \eqref{schwar2} for $c'$, differentiate and use \eqref{schwar2} again to deduce the third equation in \eqref{Einsteineq}. Together  with \eqref{schwar1} this then implies the second equation in \eqref{Einsteineq}. The first equation in \eqref{Einsteineq} follows directly from \eqref{schwar2} upon substitution of \eqref{schwar1} and its differentiated form \eqref{schwar0}.

Eliminating $f$ from the two equations \eqref{schwar1} and \eqref{schwar2}, we obtain a  differential relation between $a$ and $c$ which holds for any choice of radial coordinate:
\bee
d \ln \left(1-\frac{c^2}{\Lambda^2}\right) = -d\ln a.
\eee
Integrating gives
\bee
a=\frac{L}{1-\frac{c^2}{\Lambda^2}},
\eee
where $L$ is the value of $a$ when $c=0$, i.e., $L$ is the radius of the bolt. It has the dimension  of length and  is the only free parameter of the ES metric. It is therefore natural to rescale $c$ so that $\Lambda=L$.  Then  $c$ takes values  in  $[0,L)$ and $a$ in $[L,\infty)$, and we have the following relations which will be used frequently in the following:
\bee
\label{ac}
a=\frac{L}{1-\frac{c^2}{L^2}}, \quad \text{or} \quad  1-\frac L a = \frac{c^2}{L^2}.
\eee

A geometrically natural choice for the radial coordinate is the function $c$ itself. 
 Then $c'=1$, so that \eqref{schwar2} with $\Lambda=L$ gives
\bee
f=\frac{\frac{2 a  }{L}}{1-\frac{c^2}{L}} = \frac{2}{\left(1-\frac{c^2}{L^2}\right)^2},
\eee
and
the ES metric  takes the form
\bee
\label{ESmetriccc}
ds^2 = \frac{L^2}{\left(1-\frac{c^2}{L^2}\right)^2}(d\theta^2 + \sin^2\theta d\phi^2) + 4\frac{dc^2}{\left(1-\frac{c^2}{L^2}\right)^4}  +c^2d\tau^2.
\eee

The metric $ds^2 =4dc^2/\left(1-\frac{c^2}{L^2}\right)^4 + c^2d\tau^2 $ on the $D^2$ factor of the ES space 
 simplifies to $4dc^2 + c^2 d\tau^2$ near the origin. For the metric to be regular there, we therefore require
$\tau$ to be an angular coordinate of range $[0,4\pi)$. Defining a new angular coordinate
\bee
\chi =\frac \tau 2,
\eee
with the usual range $[0,2\pi)$, the metric on $D^2$ is 
\bee
\label{cigar}
ds^2 =4\frac{dc^2}{\left(1-\frac{c^2}{L^2}\right)^4} + 4c^2d\chi^2, \quad  c\in[0,L), \;\; \chi \in   [0,2\pi).
\eee
The finite range of $c$ is our reason for referring to this submanifold of the ES space as a disk. 

The two-dimensional geometry defined by \eqref{cigar} can be visualised as a cigar-shaped surface embedded in Euclidean three-space.  The Gauss curvature is maximal at the tip, where it equals $1/L^2$. Far away from the tip, the cigar is approximately a cylinder of radius $2L$.  Thus, we find that the single length parameter  $L$ of the ES geometry appears in three apparently distinct guises: the radius of the bolt, the curvature of the cigar at the bolt and half of the asymptotic radius of the  $O(2)$ orbits. 

The choice of $c$ as a radial coordinate and the use of the angular coordinate $\chi$ brings the ES metric into the form 
\bee
\label{ESmetricc}
ds^2 = \frac{L^2}{\left(1-\frac{c^2}{L^2}\right)^2}(d\theta^2 + \sin^2\theta d\phi^2) + 4\frac{dc^2}{\left(1-\frac{c^2}{L^2}\right)^4}  +4c^2d\chi^2,
\eee
which is 
 remarkably similar to the TN metric when similarly parametrised in terms  of the radius of its circle fibre:
 \bee
\label{TNmetricc}
ds^2_{{\text \tiny TN}}   = \frac{c^2}{\left(1-\frac{c^2}{L^2}\right)^2}(d\theta^2 + \sin^2\theta d\phi^2) +4\frac{dc^2}{\left(1-\frac{c^2}{L^2}\right)^4}  + c^2(d\psi +\cos\theta d\phi)^2.
\eee
 Here $\psi$ is the angular fibre coordinate for the TN geometry, with range $[0,4\pi)$, see \cite{AFS} for details. The TN geometry differs from the ES geometry in that its circle fibration is non-trivial, and the circles only shrink to zero size at one point (the nut, in the terminology of \cite{GH}) rather than on a two-sphere. However, both the  ES and the TN spaces contain a two-sphere's worth of  cigar-shaped submanifolds  with the  (same!) metric \eqref{cigar}. In particular, in both cases the length scale associated with  the  asymptotic radius and  the Gauss curvature at the tip is the same.

To end our discussion of the ES geometry,  we show how to switch from the radial coordinate $c$ to the more conventional Schwarzschild radial coordinate, traditionally denoted $r$.  The radial coordinate introduced by Schwarzschild (in the Lorentzian context) is  the  radius of  the $O(3)$-orbits in the ES geometry, so in terms of our notation \eqref{schwarMetric1}, $r=a$. Inserting into \eqref{ac} and solving for $c$ and $f$ gives
\bee
c=L\sqrt{1-\frac Lr}, \quad f = \frac{1}{\sqrt{1-\frac Lr}},
\eee
which result in  the familiar form of the ES metric:
\begin{equation}\label{schwarMetric2}
ds^2 =  r^2(d\theta^2 + \sin^2{\theta}d\phi^2)  + \frac{1}{V}dr^2 +4L^2Vd\chi^2, \quad V = 1 - \frac{L}{r},
\end{equation}
with the following ranges of the coordinates:
\begin{equation}
r\geq L , \quad 0 \leq \chi < 2\pi, \quad 0 \leq \theta \leq \pi, \quad 0 \leq \phi < 2\pi.
\end{equation}

\subsection{Abelian instantons on the Euclidean Schwarzschild space}

Rotationally invariant, square-integrable, closed and (anti-) self-dual two-forms on the TN and ES spaces are discussed  by Pope in \cite{Pope1} where he interprets them as the curvature of line bundles,   and studies the  Dirac operator minimally coupled to them. The two-form on the TN space is globally exact (though not $L^2$-exact), but the two-form on the ES space is essentially a Dirac monopole field and is only locally exact. Possible gauge potentials are discussed in \cite{Pope1} and  also more recently in \cite{Etesi}, but both papers consider  singular gauge potentials. We will avoid singularities in our treatment and therefore  review   the curvature  and its gauge potentials.  

In our notation,  and with the orientation defined by the volume form
\bee
e^1\wedge e^2\wedge e^3\wedge e^4 = 2a^2 f c\;\sin\theta d\theta \wedge d\phi \wedge dr \wedge d\chi,
\eee
the two-form considered in \cite{Pope1,Etesi}  is anti-self-dual, and can be written as 
\begin{align}
\label{ESF}
F= \frac{ip}{2a^2} \left(-e^1\wedge e^2 + e^3\wedge e^4\right)  = -\frac{ip}{2}\sin \theta d\theta \wedge d\phi + \frac{ipfc }{a^2} dr\wedge d\chi. 
\end{align}
The first term in the last expression  is the field of a Dirac monopole, and can be written as the exterior derivative of the usual gauge potentials 
$\frac{ip}{2} (\cos\theta \pm 1)d\phi$.  Using  \eqref{schwar1}, we can write the second term as 
\bee
\frac{ipfc }{a^2} dr\wedge d\chi = \frac{ipLa' }{a^2} dr\wedge d\chi = ip d\left(-\frac{L}{a} d\chi\right),
\eee
so that  the term in brackets is a natural candidate for a gauge potential. This is the choice adopted in \cite{Pope1}, but it is singular at the bolt, where $a=L$ but the angle $\chi$ is not defined. We obtain a regular gauge potential by adding $d\chi$, leading to the overall gauge potentials 
\begin{equation}
\label{ESNS}
A_N = i\frac{p}{2}(\cos\theta-1) d\phi +ip \left(1-\frac L a\right)d\chi , \ \ A_S = i\frac{p}{2}(\cos\theta+1)  +ip\left(1-\frac L a\right)d\chi,
\end{equation}
which are well defined on the northern  and southern hemispheres of $S^2$ respectively. They are related to each other by a  $U(1)$ gauge transformation on the overlap
\begin{equation}
A_N = A_S + \gamma^{-1} d\gamma,
\end{equation}
where $\gamma = e^{-ip\phi}$. Hence  $p$ is an integer by the Dirac quantisation condition.

Continuing our comparison with the situation on  the TN  space we note that, as a result of \eqref{ac},  the contraction of the gauge potential with the generator  $\partial_\chi$ of the $U(1)$  isometry of the ES space is
\bee
\iota_{\partial_\chi} A_N= \iota_{\partial_\chi} A_S= ip\frac {c^2} {L^2}.
\eee
This is formally also  the  function one obtains when  contracting the gauge potential  
\bee
ip\frac {c^2} {2L^2}(\cos \theta d\phi + d\psi)
\eee
 for the self-dual two-form on the TN space with the generator $2\partial_\psi$ of the $U(1)$ isometry  of TN, see \cite{AFS} for details.

\section{The gauged Dirac operator and its  zero-modes}

\subsection{The gauged Dirac operator on the  Euclidean Schwarzschild  space }

 In this section we compute the zero-modes of the Dirac operator on the ES space  coupled to the gauge potential \eqref{ESNS}.
Our results  confirm  the result of Pope in \cite{Pope1} that the space of normalisable  zero-modes has dimension $|p|^2$, but we are particularly interested in the transformation behaviour of the zero-modes under the spin group $SU(2)$ covering the isometry group $SO(3)$ of the ES geometry. The notation and technology we use is similar to that used in the analysis of the Dirac zero-modes on TN space in \cite{JS}, and designed to ease the comparison between the ES and TN situation. 

In particular, we use complex coordinates on the two-sphere obtained by the stereographic projection from the south pole, see  \cite{JS} for details on our conventions,
\begin{equation}\label{EScomplex}
z = \tan{\frac{\theta}{2}}e^{i\phi}, \ \ \bar{z} = \tan{\frac{\theta}{2}}e^{-i\phi},
\end{equation}
which are regular on a `northern patch', covering all but the south pole  of the two-sphere. We  then work with the gauge potential $A_N$ \eqref{ESNS}, which is well defined on this patch.  With the abbreviation 
\bee
q=1+\bar z z, 
\eee
one checks that 
\begin{equation}
\label{AN}
A_N = \frac{p}{2q}(zd\bar{z} - \bar{z}dz) + ip\left(1-\frac L a\right) d\chi.
\end{equation}

As in \cite{JS} we adopt the Clifford algebra convention 
\bee
\{\gamma_\alpha,\gamma_\beta\} =-2\delta_{\alpha \beta},
\eee
and the following choice of   $\gamma$-matrices: 
\begin{equation}\label{ESgamma}
\gamma^i = \begin{pmatrix}
0 & \tau_i \\
-\tau_i & 0
\end{pmatrix},
\gamma^4 = \begin{pmatrix}
0 & -i\tau_0 \\
-i\tau_0 & 0
\end{pmatrix}, 
\end{equation}
where $\tau_0$ is the $2\times2$ identity matrix and $\tau_i$, $i=1,2,3$, are the Pauli matrices.
Then 
\begin{equation}
[\gamma^i, \gamma^j] = -2i\epsilon_{ijk}\left(\begin{array}{cc}
\tau_k & 0 \\
0 & \tau_k
\end{array}\right), \ \ 
[\gamma^4, \gamma^i] = 2i\left(\begin{array}{cc}
\tau_i & 0 \\
0 & -\tau_i
\end{array}\right),
\end{equation}
so that, with the components $\omega^a_{\ b}$ given in \eqref{ESconnection}, the spin connection 
\bee
\Gamma = - \frac{1}{8}[\gamma^{\alpha},\gamma^{\beta}]\omega_{\alpha \beta}
\end{equation}
comes out as
\begin{align}
\label{singspin}
\Gamma =\! \frac{-a'}{2f}\begin{pmatrix}
i\tau_2 & 0 \\
0 & i\tau_2
\end{pmatrix}d\theta\! +\!\! \begin{pmatrix}
 -\frac{i\cos{\theta}}{2}\tau_3 + \frac{ia'\sin\theta}{2f}\tau_1 & 0 \\
0 &  -\frac{i\cos{\theta}}{2}\tau_3 +\frac{ia'\sin\theta}{2f}\tau_1
\end{pmatrix} d\phi
  +\!\frac{c'}{f}\!\begin{pmatrix}
-i\tau_3 & 0 \\
0 & i\tau_3
\end{pmatrix}d\chi.
\end{align}

We know from  \eqref{schwar2} and \eqref{ac}  that 
\bee
\frac{c'}{f}= \frac 1 2 \left(1-\frac{c^2}{L^2}\right)^2, 
\eee
which does not vanish when $c=0$. Hence  the spin connection \eqref{singspin} is singular at the bolt, where $\chi$ is not defined. As for the abelian connection discussed in the previous section, we can switch to  a regular gauge by  applying the (singular) gauge  transformation
\bee
\label{gaugetrafo}
g(\chi) =
\begin{pmatrix}
 e^{i\frac \chi 2\tau_3} & 0 \\
0 & e^{-i\frac \chi 2 \tau_3}
\end{pmatrix}.
\eee
However, it is more convenient to solve the Dirac equation  in the singular gauge \eqref{singspin} and to apply the gauge transformation  \eqref{gaugetrafo} to the solution. Since the gauge transformation satisfies
\bee
g(\chi=2\pi)= -1
\eee 
we will need to impose $\psi(\chi+2\pi) =-\psi(\chi)$  on spinors when solving the Dirac equation in the singular gauge.

With the conventions of \cite{JS} for the Dirac operator on a manifold with orthonormal frame $E_\alpha$ minimally coupled to an abelian connection $A$
\begin{align}
\label{gendirac}
\Dslash_A = \gamma^{\alpha}\iota_{E_\alpha}(d + A + \Gamma),
\end{align}
we find the following form of the Dirac operator on the ES space  minimally coupled to the connection \eqref{AN}:
\begin{equation}
\tilde{\slashed{D}}_{ES,p} = \begin{pmatrix}
0 & \tilde{T}_p^{\dagger}\\
\tilde{T}_p&0
\end{pmatrix},
\end{equation}
where
\begin{align}
\label{Diracspherical}
\tilde{T}_p &= -\begin{pmatrix}
\frac{1}{f}\partial_r + \frac{i}{2c}\partial_{\chi} - \frac{p}{2c}\left(1-\frac{L}{a}\right) + \frac{a'}{af} + \frac{c'}{2cf} & \frac{1}{a} \bar{\eth}_{\tilde s } \\
\frac{1}{a} \eth_{s}  & -\frac{1}{f}\partial_r + \frac{i}{2c}\partial_{\chi} - \frac{p}{2c} \left(1-\frac{L}{a}\right)- \frac{a'}{af} - \frac{c'}{2cf}
\end{pmatrix}, \nonumber \\
\tilde{T}^{\dagger}_p  &= \begin{pmatrix}
\frac{1}{f}\partial_r - \frac{i}{2c}\partial_{\chi} +\frac{p}{2c}\left(1-\frac L a\right) + \frac{a'}{af} + \frac{c'}{2cf} & \frac{1}{a}\bar{\eth}_{\tilde s }\\
\frac{1}{a}\eth_{s} & -\frac{1}{f}\partial_r - \frac{i}{2c}\partial_{\chi} + \frac{p}{2c}\left(1-\frac L a\right) - \frac{a'}{af} - \frac{c'}{2cf}
\end{pmatrix}.
\end{align}
Here we have used the `edth' operators  
\begin{align}
\label{ethdef}
\eth_{s} = \partial_\theta +i\frac{1}{\sin \theta } \partial_\phi
-s\frac{\cos \theta}{\sin \theta},  \quad  s= \frac 1 2 (p-1), \nonumber \\
\bar{ \eth}_{\tilde s }= \partial_\theta -i\frac{1}{\sin \theta } \partial_\phi
+ \tilde  s\frac{\cos\theta}{\sin \theta}, \quad \tilde s =\frac 1 2 (p+1),
\end{align}
introduced by Newman and Penrose   \cite{NP} and frequently used to write the Dirac operator on the two-sphere, see for example \cite{Dray2}.
As explained in \cite{JS}, they 
the are related to  operators in terms of our complex coordinate $z$  via
\begin{align}
(q\bar{\partial_z} + sz)e^{is\phi} = e^{i\tilde s \phi}\eth_s  \quad \text{and} \quad (q\partial_z -\tilde s\bar{z})e^{i\tilde s\phi} &= e^{is \phi}\bar{\eth}_{\tilde s }.
\end{align}
 It follows that, with the gauge change 
\bee
G(\phi) = \text{diag}( e^{i s \phi},   e^{i\tilde s\phi}    ,e^{is \phi}, e^{i\tilde s\phi}),
\eee
  the operator 
\bee
\label{ESA}
\slashed{D}_{ES,p}  = G(\phi)\tilde{\slashed{D}}_{ES,p} G^{-1}(\phi)= 
\begin{pmatrix}
0 & T_p^{\dagger}\\
T_p&0
\end{pmatrix},
\eee
which we will study for the remainder of this section, has  the simpler form
\begin{align}
T_p &= -\begin{pmatrix}
\frac{1}{f}\partial_r + \frac{i}{2c}\partial_{\chi} - \frac{p}{2c}\left(1-\frac{L}{a}\right) + \frac{a'}{af} + \frac{c'}{2cf} & \frac{1}{a}\left(q\partial_z - \frac{1}{2}(p+1)\bar{z}\right) \\
\frac{1}{a}\left(q\bar{\partial}_z + \frac{1}{2}(p-1)z\right) & -\frac{1}{f}\partial_r + \frac{i}{2c}\partial_{\chi} - \frac{p}{2c}\left(1-\frac{L}{a}\right)- \frac{a'}{af} - \frac{c'}{2cf}
\end{pmatrix}, \nonumber \\
T^{\dagger}_p &= \begin{pmatrix}
\frac{1}{f}\partial_r - \frac{i}{2c}\partial_{\chi} + \frac{p}{2c}\left(1-\frac{L}{a}\right)  + \frac{a'}{af} + \frac{c'}{2cf} & \frac{1}{a}\left(q\partial_z - \frac{1}{2}(p+1)\bar{z}\right) \\
\frac{1}{a}\left(q\bar{\partial}_z + \frac{1}{2}(p-1)z\right) & -\frac{1}{f}\partial_r - \frac{i}{2c}\partial_{\chi} + \frac{p}{2c}\left(1-\frac{L}{a}\right)  - \frac{a'}{af} - \frac{c'}{2cf}
\end{pmatrix}.
\end{align}  

Note that the $z$-dependence of $ \slashed{D}_{ES,p}$  only occurs in the off-diagonal part of both  $T_p$ and $T^\dagger_p$ as  
 \bee
 \begin{pmatrix}  0 & q\partial_z - \frac{1}{2}(p+1)\bar{z}\ \\
 q\bar{\partial}_z + \frac{1}{2}(p-1)z     &  0 
    \end{pmatrix}.
 \eee
This is the Dirac operator on the two-sphere twisted by a Dirac monopole, and  was studied
 in   \cite{JS2} using the same complex coordinate $z$ used here. As explained there and reviewed in the Appendix, 
the spin bundle over the two-sphere twisted by the Dirac  monopole of charge $p$  is  a direct sum of line bundles $H^{p-1}\oplus H^{p+1}$, where $H$ is  the hyperplane bundle $H$ over  $S^2\simeq\mathbb{CP}^1$. The operator $q\bar{\partial}_z + \frac{1}{2}(p-1)z$ maps local sections  of $H^{p-1}$  to local sections of  $H^{p+1}$, while   the operator $q\partial_z - \frac{1}{2}(p+1)\bar{z} $  maps local  sections of  $H^{p+1}$ to local sections of $H^{p-1}$.
It follows from 
\begin{equation}\label{ESdd}
 q\bar{\partial}_z + \frac{1}{2}(p-1)z = q^{\frac{1}{2}(-p+3)}\bar{\partial_z}q^{\frac{1}{2}(p-1)}, \ \ q\partial_z - \frac{1}{2}(p+1)\bar{z} = q^{\frac  12 (p+3)}\partial_zq^{-\frac 1 2 (p+1)},
\end{equation}
that these operators  annihilate local sections of the form $q^{\frac1 2 (1-p)} p_1(z)$ and $q^{\frac 1 2 (1+p)} p_2(\bar{z})$, with  regularity at $z=0$ requiring that  $p_1(z)$ is a polynomial of degree $p-1$  and $p_2(\bar{z})$ to be  a polynomial of degree $-p-1$.
Clearly, the former is only possible for $p\geq 1$ and the latter if  $p \leq -1$. 

\subsection{Zero-modes in complex coordinates}
In order to solve for the zero-modes of the Dirac operator, it is best to rewrite it using the identities \eqref{schwar1}, \eqref{schwar2} and \eqref{ac}.  In particular, we use
\bee
\frac{f}{2c}\left(1-\frac La \right) =\frac{fc}{2L^2} =\frac{a'}{2L}, 
\eee
and 
\begin{align}
\frac{f}{2c} =\frac{f}{2c}\left(1-\frac La \right) + \frac{Lf}{2ac} 
=\frac{a'}{2L} + \frac{c'}{c}\frac{1}{1-\frac{c^2}{L^2}}  = \frac{a'} {2L} +\frac 1 2 \frac{d}{dr} \ln (ac^2).
\end{align}
The ansatz
\bee
\Psi = \begin{pmatrix} 0 \\0 \\ R(r)e^{-i(n+\frac 1 2 )\chi} q^{\frac 1 2 (1-p)} \sum_{k=0}^{p-1}a_kz^k\\0 \end{pmatrix},
 \qquad p \geq 1
\eee
for $\slashed{D}_{ES,p} \Psi=0$,  then leads to the  radial equation
\bee
\frac{d\ln R}{dr} + \frac{d \ln a \sqrt{c}} {dr}   =  \left(n+\frac 1 2\right) \frac {d \ln c \sqrt{a}}{dr} +\frac{n+ \frac  1 2 -p}{2L}\frac{da}{dr},
\eee
which can be integrated to 
\bee
R=Kc^na^{\frac n 2 -\frac 34} e^{(-p+n+\frac 1 2 )\frac{a}{2L}},
\eee
where $K$ is an arbitrary constant.
Note that this expression was obtained without choosing any particular radial coordinate $r$.
The function $R$  is square-integrable  with respect to the measure $ca^2 fdr$ if the exponential factor is decaying for large $a$ and if the (integer) power of $c$ is non-negative, i.e., if
\bee
0\leq  n \leq p-1.
\eee

Similarly, the ansatz
\bee
\Psi = \begin{pmatrix} 0 \\0 \\0 \\R(r)e^{i(n+\frac 1 2 )\chi} q^{\frac 1 2 (1+p)} \sum_{k=0}^{-p-1}a_k\bar{z}^k\end{pmatrix}, 
\qquad p \leq  -1, 
\eee
leads to the radial equation 
\bee
\frac{d \ln R}{dr} + \frac{d \ln a \sqrt{c}} {dr}  =  \left(n+\frac 1 2\right) \frac {d \ln c \sqrt{a}}{dr} +\frac{n+ \frac  1 2 +p}{2L}\frac{da}{dr}.
\eee
It is solved by 
\bee
R=Kc^na^{\frac n 2 -\frac 34} e^{(p+n+\frac 1 2 )\frac{a}{2L}},
\eee
which is square-integrable if 
\bee
0 \leq n \leq -p -1.
\eee
It is easy to check that a similar ansatz with non-vanishing entries in the first two components of the spinor $\Psi$ does not produce solutions with the behaviour at the bolt and at infinity  required for square-integrability.

Using \eqref{gaugetrafo} to switch to the regular gauge, we thus   have  the following general form 
\bee
\label{zeroplus}
\Psi_{\text{\tiny reg}} = \begin{pmatrix} 0 \\0 \\  e^ {-in\chi} c^n a^{\frac n 2 -\frac 34} e^{(-p+n+\frac 1 2 )\frac{a}{2L}}
q^{\frac 1 2 (1-p)} \sum_{k=0}^{p-1}a_kz^k \\ 0 \end{pmatrix}, 
 \qquad p \geq 1, \quad 0\leq n \leq p-1,
\eee
of zero-modes for positive $p$, and 
\bee
\label{zerominus}
\\Psi_{\text{\tiny reg}}  = \begin{pmatrix} 0 \\0 \\0 \\  e^{in\chi} c^na^{\frac n 2 -\frac 34} e^{(p+n+\frac 1 2 )\frac{a}{2L}}q^{\frac 1 2 (1+p)} \sum_{k=0}^{-p-1}a_k\bar{z}^k\end{pmatrix}, 
\qquad p \leq  -1,  \quad 0\leq n \leq -p-1,
\eee
for negative $p$.

The expressions \eqref{zeroplus} and \eqref{zerominus} give the zero-modes entirely in terms of geometrical data associated to the ES space. They warrant several comments.

The functions $q^{\frac 1 2 (1-p)} z^k$ and  $q^{\frac 1 2 (1+p)}\bar{z}^k$, $k=0,\ldots |p|+1$, expressing the $S^2$-dependence of the zero-modes
 form  irreducible representations of $SU(2)$ of spin $j = (p-1)/2$ for $p\geq 1$ and $j=(-p-1)/2$ for $p\leq 1$.
This is  shown in \cite{JS} and  reviewed in the Appendix, see particularly  \eqref{yjm} and \eqref{y-jm}. 
  Hence, the space of the above zero-modes splits into $|p|$ copies of $|p|$-dimensional irreducible representations of $SU(2)$.
This reproduces the dimension  $|p|^2$ computed by Pope, and exhibits the transformation behaviour under $SU(2)$.

Note also that the  zero-modes are holomorpic in the  coordinate 
\bee 
w= ce^{-i\chi}
\eee
  when $p$ is positive, and anti-holomorphic in $w$ when $p$ is negative. This echoes a similar observation for the TN manifold. In both cases, the zero-modes are particularly simple when written in terms of complex   coordinates for the cigar-shaped submanifolds. 

Finally, we express the zero-modes in terms of the Schwarzschild coordinate $r$. They then take the form
\bee
\label{zeroplusr}
 \Psi_{\text{\tiny reg}}  = \begin{pmatrix} 0 \\0 \\  e^ {-in\chi} (r-L)^{\frac n 2}  r^{-\frac 34} e^{(-p+n+\frac 1 2 )\frac{r}{2L}}
q^{\frac 1 2 (1-p)} \sum_{k=0}^{p-1}a_kz^k \\ 0 \end{pmatrix}, 
 \qquad p \geq 1, \quad 0\leq n \leq p-1,
\eee
of zero-modes for positive $p$, and 
\bee
\label{zerominusr}
 \Psi_{\text{\tiny reg}}  = \begin{pmatrix} 0 \\0 \\0 \\  e^{in\chi} (r-L)^{\frac n 2}r^{ -\frac 34} e^{(p+n+\frac 1 2 )\frac{r}{2L}}q^{\frac 1 2 (1+p)} \sum_{k=0}^{-p-1}a_k\bar{z}^k\end{pmatrix}, 
\qquad p \leq  -1,  \quad 0\leq n \leq -p-1,
\eee
for negative $p$.

Focusing on the case $p\geq 1$ and using $k = m + j$, where $j = (p-1)/2$, the functional dependance of the zero-mode \eqref{zeroplusr} can be written as
\begin{align}
\psi(r,z,\tau)  &=e^ {-in\chi} (r-L)^{\frac n 2}  r^{-\frac 34} e^{(-p+n+\frac 1 2 )\frac{r}{2L}}
q^{-j}\sum_{m=-j}^{j}a_mz^{m+j} \nonumber \\
&= e^ {-in\chi} (r-L)^{\frac n 2}  r^{-\frac 34} e^{(-p+n+\frac 1 2 )\frac{r}{2L}}
\sum_{m=-j}^{j}a_m\left(\cos\frac{\theta}{2}\right)^{j-m}\left(\sin\frac{\theta}{2}\right)^{j+m}e^{i(j+m)\phi}, 
\end{align}
where we have used \eqref{EScomplex}.  Fixing a  value of $m$ and neglecting overall factors, we obtain the typical  probability distribution
\begin{equation}
|\psi|^2(r,z,\tau) =  (r-L)^{ n}  r^{-\frac 32} e^{(-p+n+\frac 1 2 )\frac{r}{L}}
(r-x_3)^{j+m}(r+x_3)^{j-m},
\end{equation}
where $(x_1,x_2,x_3) = (r\sin\theta\cos\phi, r\sin\theta\sin\phi, r\cos\theta)$, which resembles the distribution of the zero-modes of  the Dirac operator on TN plotted  in \cite{JS}.
Note that, in the Schwarzschild coordinates,  the radial  volume element used to normalise the probability distributions is the same as in $\mathbb{R}^3$,  i.e., $cfa^2dr = r^2dr$.

\begin{figure}[!ht]
\begin{centering}
\includegraphics[width=7.5truecm]{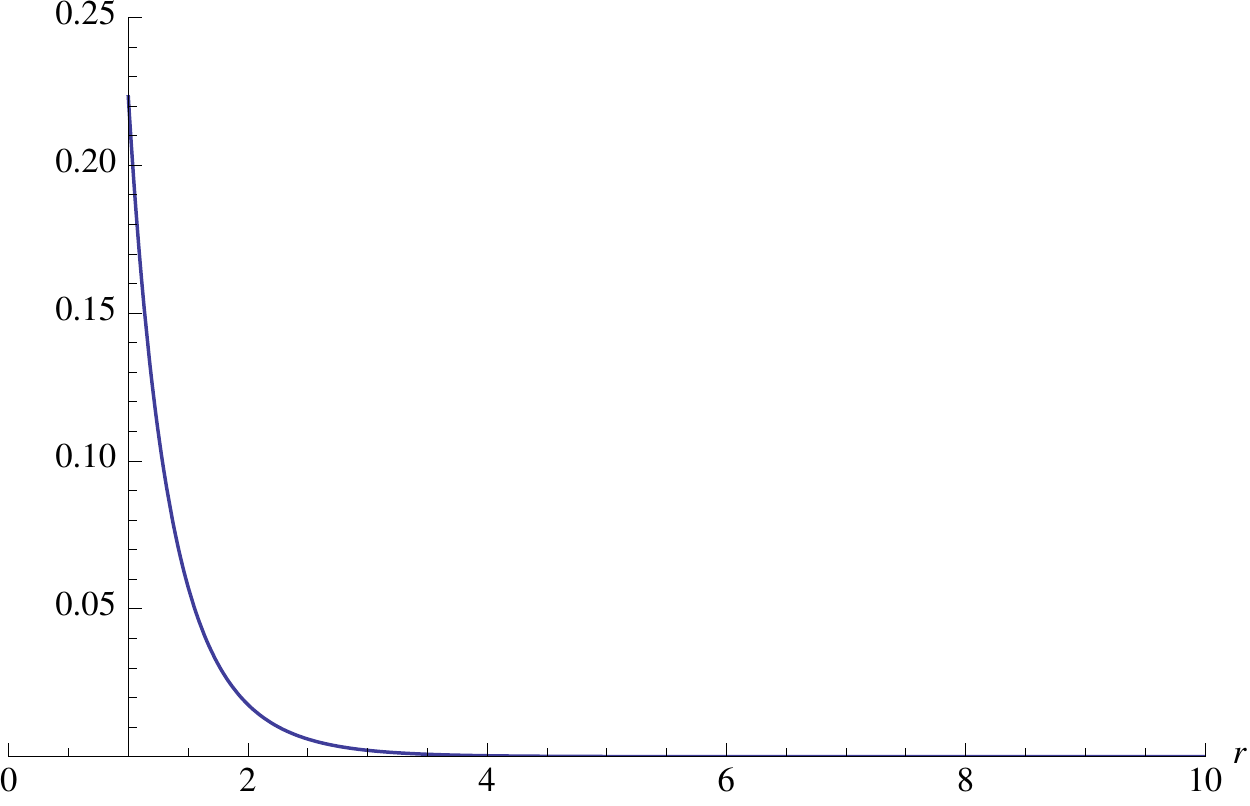}
\includegraphics[width=7.5truecm]{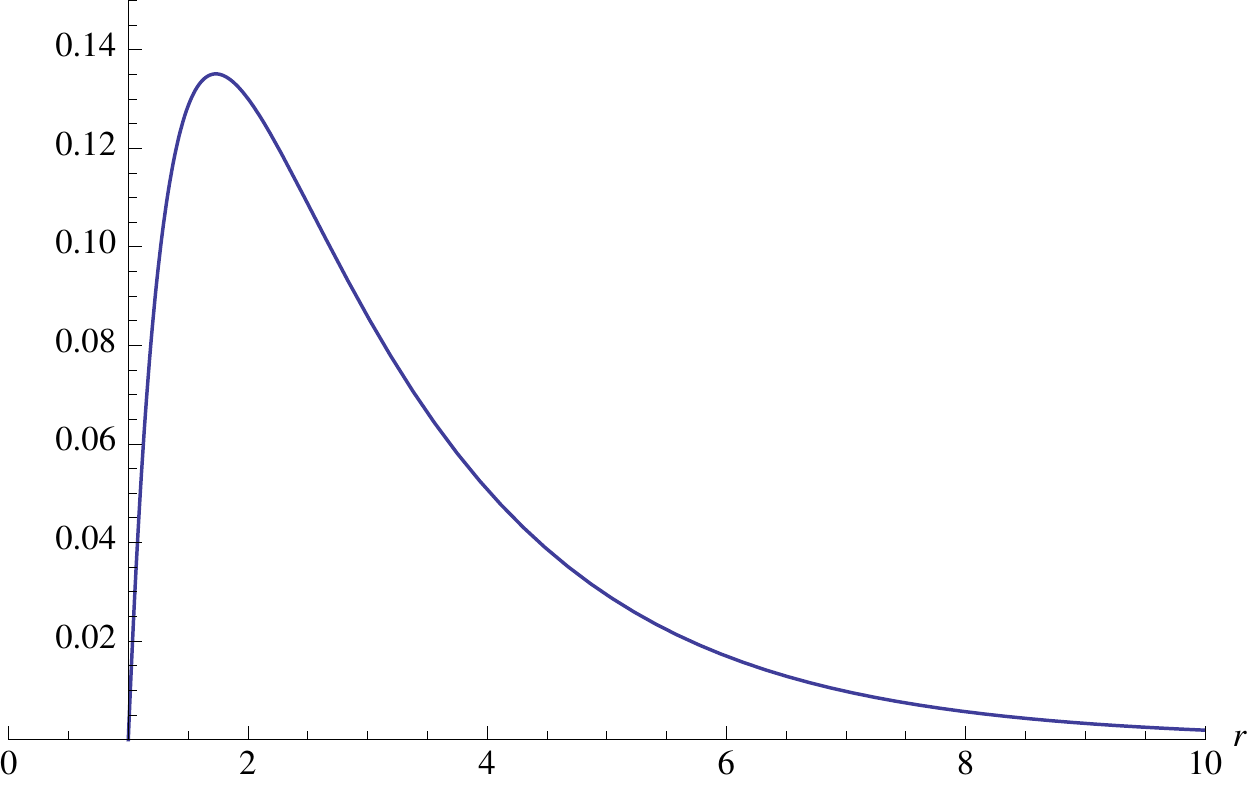}
\caption{Plot of the $r$-dependence of $|\psi^0_{\frac{1}{2}}|^2$ (left) and $|\psi^1_{\frac{1}{2}}|^2$ (right) with $L=1$.}
\label{rdependence}
\end{centering}
\end{figure}

Finally, we focus on the  case of spin-$\frac{1}{2}$ states,  i.e.,  $j = \frac{1}{2}$ which, assuming for simplicity $p\geq 0$, are obtained from \eqref{zeroplusr} by picking $p=2$. In this case we have $n = 0,1$ and  obtain two spin-$\frac{1}{2}$ doublets
\begin{equation}
\psi^0_{\frac{1}{2}}=  r^{-\frac 34} e^{-\frac{3}{4L}r }q^{-\frac{1}{2}}(a_0 + a_1z), \quad \psi^1_{\frac{1}{2}}=
e^ {-i\chi} (r-L)^{\frac 1 2}  r^{-\frac 34} e^{-\frac{1}{4L}r}q^{-\frac{1}{2}}(a_0 + a_1z).
\end{equation}
Both states are exponentially localised at the bolt. At  the bolt ($r=L$),  $\psi^0_{\frac  12 }$
 has a finite value but  $\psi^1_{\frac  12 }$  vanishes. We  plot  the  $r$-dependence for  the squared norm  of both zero-modes in  Fig.~ \ref{rdependence}.

\section{The gauged scalar Laplace operator and its bound states}
\subsection{Dirac Laplacian versus scalar Laplacian}

 The general Lichnerowicz-Weitzenb\"ock formula for a Dirac operator on a Riemannian manifold twisted   by a $U(1)$ gauge potential $A$ (see for example \cite{Jost}) is
\begin{equation}\label{DDRA}
\slashed{D}_A^{\dagger}\slashed{D}_A = -g^{\mu\nu}D_{\mu}D_{\nu} + \frac{1}{4}R + \frac{1}{2}[\gamma^\alpha,\gamma^\beta]F_{\alpha\beta},
\end{equation} 
where $F$ is the curvature of $A$ and $D_{\mu} = \partial_{\mu} - \Gamma_{\mu} + A_{\mu}$. Using this, and taking into account that the scalar curvature $R$ of the ES geometry is zero, we see that, for a scalar $\phi$ and a spinor $\Psi$, 
\begin{equation}\label{ESDL}
\slashed{D}^{\dagger}_A\slashed{D}_A\phi\Psi = -(g^{\mu\nu}D_{\mu}\partial_{\nu}\phi)\Psi - (g^{\mu\nu}D_{\mu}D_{\nu}\Psi)\phi + \frac{1}{2}[\gamma^\alpha,\gamma^\beta]F_{\alpha\beta}\phi\Psi.
\end{equation}
For vanishing scalar curvature $R$ and (anti-) self-dual curvature $F$, the presence of covariantly constant spinor would allow one to relate  the spectrum of the Dirac Laplacian $\slashed{D}^{\dagger}_A\slashed{D}_A$ to the  gauged scalar Laplacian 
\begin{equation}\label{ESlaplace}
\Delta_A = \frac{1}{\sqrt{g}}(\partial_{\mu}+A_\mu)g^{\mu\nu}\sqrt{g}(\partial_{\nu}+A_\nu).
\end{equation}
This was the case  for  TN \cite{JS,JS2},  but  the  ES space  does not admit covariantly constant spinors. In this paper we therefore only look at the spectrum of the gauged  scalar Laplacian.

\subsection{The gauged scalar Laplace operator on the Euclidean Schwarzschild space}

The Laplace  operator \eqref{ESlaplace} associated to the generalised metric \eqref{schwarMetric1} is
\begin{equation}
\Delta_{ES} = \frac{1}{a^2cf}\partial_r\left(\frac{a^2c}{f}\right)\partial_r + \frac{1}{4c^2}\partial_{\chi}^2 + \frac{1}{a^2}\Delta_{S^2},
\end{equation}
where  $\Delta_{S^2}$ is the Laplace operator on the two-sphere of unit radius:
\begin{align}
\Delta_{S^2} = \frac{1}{\sin^2\theta}\partial_{\phi}^2 + \frac{1}{\sin\theta}\partial_{\theta}\sin\theta\partial_{\theta}  =  q^2\partial_z\bar \partial_z.
\end{align}

Minimally coupling the Laplace operator to the potential \eqref{AN} yields
\begin{equation}\label{ESlaplace2}
\Delta_{ES,p} = \frac{1}{a^2cf}\partial_r\left(\frac{a^2c}{f}\right)\partial_r + \frac{1}{4c^2}\left(\partial_{\chi} +ip\left(1-\frac La\right)\right)^2 + \frac{1}{a^2}\Delta_{S^2,p} \, ,
\end{equation}
where now $\Delta_{S^2,p} $ is the  Laplace operator on the two-sphere of unit radius minimally coupled to the potential \eqref{ESNS} which, on the two-sphere, is the potential of the Dirac monopole. The Laplace operator on the two-sphere coupled to the Dirac monopole is a  much studied operator, and its eigensections, often called monopole spherical harmonics, are well know. The non-triviality of the monopole bundle means that one has to work in patches, and many different conventions exist in the literature, using different coordinates on the two-sphere and different coverings. We will work with the complex stereographic coordinate $z$ already used in the discussion of the Dirac operator, which covers all of the two-sphere except the south pole.
For details we refer to the Appendix \ref{s2app}, where it is shown that 
 \bee
\Delta_{S^2,p}  = \frac{1}{2}\left( q^{\frac{p}{2}+2}\partial_z q^{-p}\bar \partial_z q^{\frac{p}{2}} +  
  q^{-\frac{p}{2}+2}\bar \partial_z q^p \partial_z q^{-\frac{p}{2}} \right).
\eee
Local eigensections of this operator are given in terms of complex coordinates in \eqref{mopoharmonics} in the  Appendix, where they are denoted  as follows:
 \bee
 \label{jCond}
  y^j_{\frac{p}{2}m}, \quad j   = \frac{|p|}{2} + \NN^0, \qquad m=-j,-j+1,\ldots,j-1,j.
  \eee 
  They satisfy
  \bee
  \Delta_{S^2,p} y^j_{\frac{p}{2}m}= \left(-j(j+1) + \frac{p^2}{4}\right)y^j_{\frac{p}{2}m}.
  \eee

When looking  for solutions of 
\begin{equation}
 -\Delta_{ES,p}\phi = \lambda\phi
\end{equation}
we therefore assume  the factorised form 
\begin{equation}
\label{ansatz}
\phi= R(r)e^{in\chi}y^j_{\frac{p}{2}m}, 
\end{equation} 
 where
\bee
n\in \ZZ,
\eee
since the range of $\chi $ is $[0,2\pi)$.
This ansatz yields the following radial equation: 
\begin{align}
\label{eigenLambda}
V\frac{d^2R(r)}{d r^2} + \left(\frac{L}{r^2} + \frac{2V}{r}\right)\frac{d R(r)}{d r} -\frac{1}{V}\left(\frac{n}{2L} +\frac{p}{2}\left(\frac 1 L -\frac 1 r\right)\right)^2R(r) \qquad \qquad  \qquad \qquad  \hfill  \nonumber \\
 + \frac{1}{r^2}\left(-j(j+1) + \frac{p^2}{4}\right)R(r) + \lambda R(r) = 0.
\end{align}

In order to facilitate comparison with previous analysis of the spectrum of the ES geometry in the ungauged case \cite{Fawcett}, we introduce the  dimensionless coordinate 
\begin{equation}
x = \frac{r}{L},
\end{equation}
and the  so-called tortoise coordinate 
\begin{equation}
x^* = x + \ln{(x-1)}.
\end{equation}
 In terms  of the  rescaled eigenvalue $\bar{\lambda} = L^2\lambda$, the radial eigenvalue  equation \eqref{eigenLambda} then  reads
\begin{equation}
\frac{x-1}{x}\frac{d^2R}{dx^2} + \left(\frac{2}{x}-\frac{1}{x^2}\right)\frac{dR}{dx} + \left[- \frac{x}{x-1}\left(\frac{n}{2}+\frac{p}{2}\left(1-\frac 1 x\right)\right)^2 - \frac{j(j+1)}{x^2} + \frac{p^2}{4x^2} + \bar{\lambda}\right]R = 0.
\end{equation}
Setting 
\begin{equation}
f(x)= xR(xL),
\end{equation} 
we obtain
\begin{equation}
\frac{1}{x^3}\frac{df}{dx} + \left(\frac{1}{x} - \frac{1}{x^2}\right)\frac{d^2f}{dx^2} - \frac{f}{x^4} + \left[- \frac{1}{x-1}\left(\frac{n}{2}+\frac{p}{2}\left(1-\frac 1 x\right)
\right)^2 - \frac{j(j+1)}{x^3} + \frac{p^2}{4x^3} + \frac{\bar{\lambda}}{x}\right]f = 0.
\end{equation}
Finally noting that 
\begin{equation}
\frac{1}{x^3}\frac{df}{dx} + \left(\frac{1}{x} - \frac{1}{x^2}\right)\frac{d^2f}{dx^2} = \frac{1}{x-1}\frac{d^2f}{d{x^*}^2},
\end{equation}
and multiplying by $x-1$, the above equation can be recast as
\begin{equation}
\frac{d^2f}{d{x^*}^2} = \left[\left(\frac{n}{2}+\frac{p}{2}\left(1-\frac 1 x\right)\right)^2 - \left(1-\frac{1}{x}\right)\left(-\frac{j(j+1)}{x^2} + \frac{p^2}{4x^2} - \frac{1}{x^3} + \bar{\lambda}\right)\right]f,
\end{equation}
which, in the case $p=0$, reduces to the eigenvalue equation discussed in \cite{Fawcett}.

We have not been able to solve the eigenvalue problem  \eqref{eigenLambda} exactly. However, we can bring it into a standard from and then  solve it numerically by factoring  
\begin{equation}
R = \frac{g(r)}{r\sqrt{V}}.
\end{equation}
This turns \eqref{eigenLambda} into the  Sturm-Liouville problem  
\begin{equation}\label{sturm}
-g'' + Qg = \frac{\lambda}{V}  g, \qquad r\in [L,\infty),
\end{equation}
with 
\begin{equation}\label{ESpotential}
Q(r) = -\frac{L^2}{4r^4V^2} + \frac{1}{V^2}\left(\frac{n+p}{2L} - \frac{p}{2r}\right)^2+ \frac{1}{Vr^2}\left[j(j+1) -\frac{p^2}{4}\right].
\end{equation}
 This Sturm-Liouville problem  can be solved numerically even though the function 
 \bee
 1/V=r/(r-L)
 \eee 
 appearing on the right hand side is singular at $r = L$. 

Before turning to the numerical calculations,  we can gain an  intuitive understanding  of  why the potential \eqref{ESpotential} supports bound states by considering  the Sturm-Liouville problem \eqref{sturm} in the limit of large $r$. Rather surprisingly,  it then takes the form  of an eigenvalue equation for the gauged Laplace operator on the TN space, which can be solved exactly and which was shown in \cite{JS2} to have infinitely many bound states.

To obtain the form in the limit $r\to \infty$, we use 
\begin{equation}
\frac{1}{V} =  1 + \frac{L}{r} + \frac{L^2}{r^2} + \ldots .
\end{equation}
Inserting this into $Q(r)$, and keeping terms up to order $r^{-2}$ we get
\begin{align}
\label{ESinfinity}
Q(r)  \approx  \frac{(n+p)^2}{4L^2} + \frac{ n  (n+p)}{2Lr} + \left(\frac {(n+p)(3n-p)}{4} +j(j+1)\right)\frac{1}{r^2}. 
\end{align}
We can see that in the case $p\geq 0$ the Coulomb term is attractive if $
-p < n  <  0$ 
while for $p < 0$ it is attractive if 
$0 < n  < -p$.
Combining these, the condition for an attractive Coulomb potential is
\begin{equation}
\label{ESBSC}
0<n^2< p^2.
\end{equation}


Replacing the function $Q$ appearing in \eqref{sturm} by its asymptotic from  \eqref{ESinfinity} and  approximating $1/V\approx 1+\frac L r$  yields the Sturm-Liouville equation 
\begin{equation}
\label{asyform}
-g'' + \left(\frac{(n+p)^2}{4L^2} +  \frac{n  (n+p)}{2Lr} + \left(\frac {(n+p)(3n-p)}{4} +j(j+1)\right)\frac{1}{r^2}\right) g =  \lambda \left( 1+\frac L r\right) g,
\end{equation}
which is of the same form as  the radial equation appearing in the study of the gauged Laplace operator on the TN space in \cite{JS2}.  That equation  also involves parameters, called $s,p$ and $j$  in \cite{JS2}. Re-naming them $s, \tilde p$ and $ \tilde j$ to avoid confusion with the ES parameters, the equation reads
\begin{equation}
\label{radialSch}
\left[-\frac{1}{r^2}\partial_r(r^2\partial_r) + \frac{\tilde j(\tilde j+1)}{r^2} + \left(\frac{2s^2}{L} - \frac{\tilde ps}{L} - E L\right)\frac{1}{r} + \left(\left(\frac{s-\frac{\tilde p}{2}}{L}\right)^2 - E\right)\right]R(r) = 0.
\end{equation}
In terms of $u(r)=r R(r)$, it takes the form
\begin{equation}
-u'' + \left(\frac{(2s-\tilde p)^2}{4L^2} + \frac{2s\left(2 s-\tilde p \right) }{2Lr} +\tilde j(\tilde j+1) \frac{1}{r^2} \right)u = E\left(1+\frac L r\right)u.
\end{equation} 
This agrees with \eqref{asyform}
provided we identify
\begin{equation}
\label{sub1}
\tilde p = p, \quad 2s=-n \quad \text{or} \quad \tilde p = -p, \quad 2s =n,
\end{equation}
and 
\begin{equation}
\label{sub2}
\tilde j (\tilde j +1 ) = \frac {(n+p)(3n-p)}{4}+j(j+1).
\end{equation}
The TN spectrum can be computed exactly \cite{JS2}, and is given by 
\begin{equation}
\label{Espectrum}
E(\tilde p,s,N) = \frac{2}{L^2}\left(  s^2 - \frac{ \tilde ps}{2}  +  N\sqrt{N^2 - s^2+\frac{\tilde p^2}{4}} - N^2\right),
\quad N  =  |s|+1, |s|+2,  \ldots .
\end{equation}
With the substitution \eqref{sub1},  this  gives the formula 
\bee
\label{lambdaapprox}
 E\left(p,-\frac{n}{2},N\right)=
 \frac{2}{L^2}\left(  \frac{n^2}{4} + \frac{ pn}{4}  +  N\sqrt{N^2 - \frac{n^2}{4}+ \frac{p^2}{4 }} - N^2\right), \;\; N=\frac{|n|}{2}+1, \frac{|n|}{2}+2,\ldots .
\eee
 for approximate eigenvalues for \eqref{sturm}.  Note that it does not matter  which of the two substitutions in \eqref{sub1} we use. Note also that the value of $j$ does not enter the approximation since the TN eigenvalues do not depend on it (though it does restrict the range of allowed values for $\tilde j$).  We will compare the approximations \eqref{lambdaapprox}  with numerical solutions of the eigenvalue problem \eqref{sturm} in the next section.

 Before leaving the TN problem, we note that  the necessary condition \eqref{ESBSC} for  bound states in the gauged ES problem has a close analogue in the condition $4 \tilde s^2 <  \tilde p^2, $
 which was shown in \cite{JS2} to be necessary for bound states in the gauged TN problem. This condition can be understood   in terms of the binding  due to the magnetic field restricted to the cigar-shaped submanifold with the metric \eqref{cigar}, as  discussed in detail in \cite{JS2}. Even though that discussion takes place entirely in the context of the TN geometry, it applies to the ES space, too, because of the similarities in the magnetic fields and the cigar geometries in the two spaces.

\subsection{Numerical computation of the spectrum}

The numerical solution of the  eigenvalue problem \eqref{sturm} in the interval $(L,\infty)$ requires information on the nature of the  endpoints,  in particular whether they are limit-point or limit-circle.  An endpoint is a limit-point  for an eigenvalue $\lambda$ if there exist two linearly independent solutions of which one is locally square-integrable near that endpoint and the other is not \cite{RS}. We now show that  for   \eqref{sturm}, both $r=L$ and $r=\infty$ are   limit-points.

For the endpoint $r=L$ and a given $\lambda\in\mathbb{C}$,   we therefore look for  independent  solutions  $g(r,\lambda)$ of \eqref{sturm} such that 
\begin{equation}
\label{limitpoint}
\int_L^{L+\delta} \frac{|g(r,\lambda)|^2}{V(r)} dr, \quad \delta < \infty, 
\end{equation}
diverges for one solution and converges for another. 
Introducing the variable $ 
\rho= r-L$
 we observe that, for small $\rho$, we have 
 $V \approx  \rho/L$.
Keeping only leading terms in   equation \eqref{sturm}, we  obtain
\begin{equation}
\label{ESL}
g'' - \frac{n^2-1}{4\rho^2}g = 0.
\end{equation}
Assuming solutions of the form $g = \rho^{\beta}$ we find that $\beta = \frac{1}{2}\pm \frac n 2$ and hence
the solutions 
\begin{equation}
g^L_\pm(r) = A_{\pm}\rho^{\frac 1 2\pm \frac n 2}, \ A_{\pm} \ {\rm constants}. 
\end{equation}
For  these solutions the potentially most divergent term  in the integral \eqref{limitpoint} is
\begin{align}
\int_0^\delta \frac L \rho |g^L_\pm|^2  d\rho \propto  \int_0^{\delta}\rho^{\pm n}d\rho.
\end{align}
The  integral for precisely one of the signs (and hence one of the solutions)  diverges provided $|n| \geq 1$.
Since $n$ is an integer, this is automatically satisfied when the condition
  \eqref{ESBSC} for an attractive Coulomb potential is satisfied. Thus $r = L$ is a limit-point.

Turning to  the second endpoint $r=\infty$, we note  from \eqref{ESinfinity}  that,  in this limit,  $V\approx 1$. Thus,  keeping only   leading terms, the eigenvalue equation \eqref{sturm} reduces to
\begin{equation}
g'' + \left(\lambda-\frac{(n+p)^2}{4L^2}\right)g = 0, 
\end{equation}
which is solved by 
\begin{equation}
g^\infty_\pm(r) = A_{\pm}e^{\pm\sqrt{\frac{(n+p)^2}{4 L^2} - \lambda}\, r}, \ A_{\pm} \ {\rm constants}.
\end{equation}
Using again $V \approx 1 $ in this limit,  we see that  the integral 
\bee
\int_\delta^\infty |g^\infty_\pm(r)|^2dr, \quad \delta < \infty, 
\eee
 converges for one solution and diverges for the other if and only if
\begin{equation}\label{condThree}
\lambda < \frac{(n+p)^2}{4L^2}. 
\end{equation}
This condition is analogous to the condition 
\bee
E < \frac{(2s-\tilde p)^2}{4L^2}
\eee
for the  energy of bound states in  the gauged TN  equation \eqref{radialSch}, see  \cite{JS2}  for a detailed discussion.

\begin{figure}[!h]
\begin{centering}
\includegraphics[width=6.4truecm]{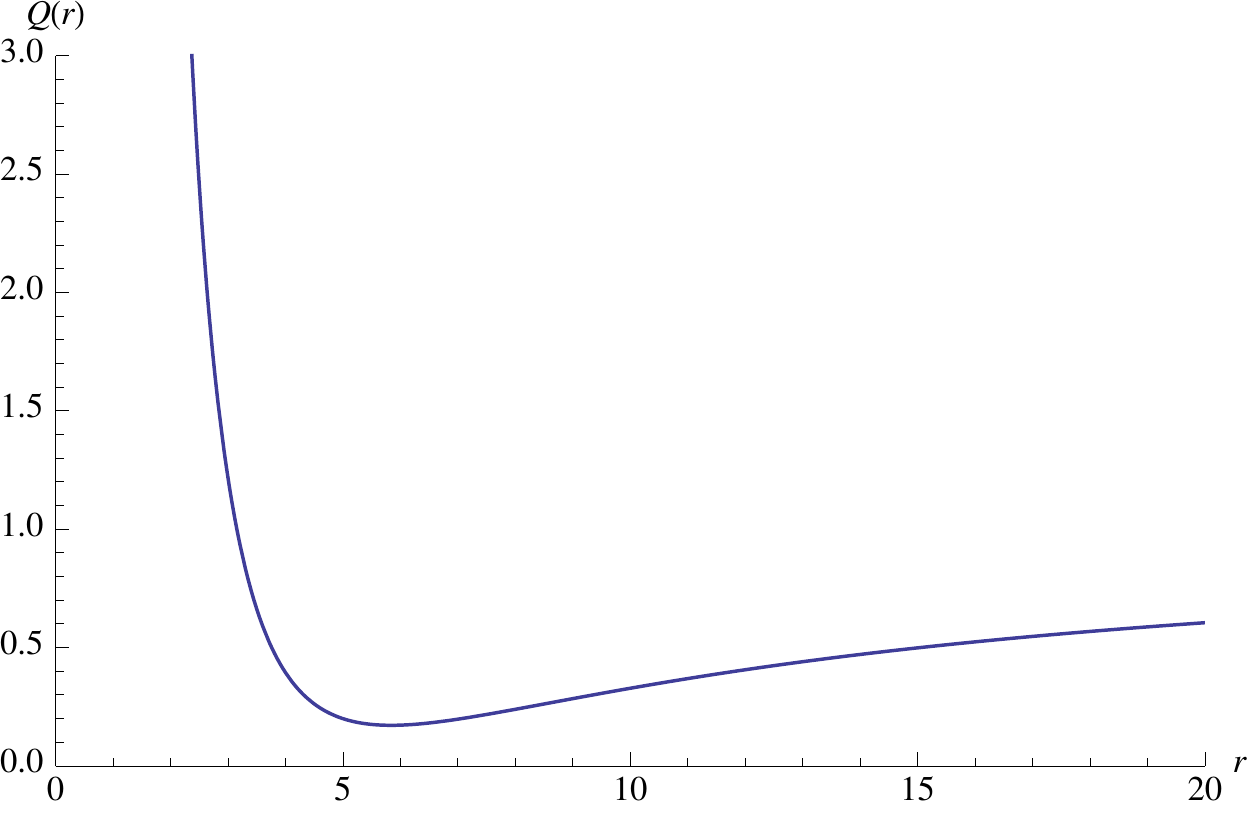} \quad  \quad 
\includegraphics[width=6.4truecm]{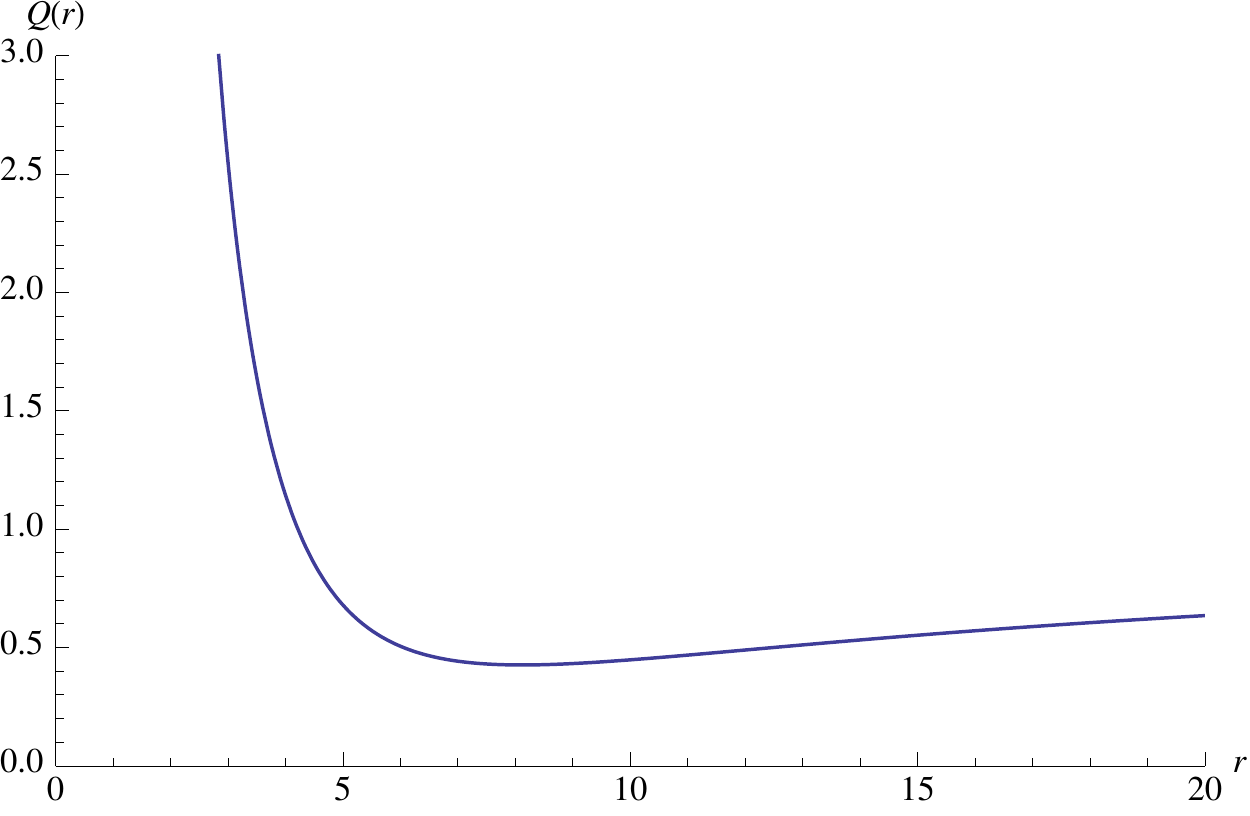}
\caption{Plot of the effective potential \eqref{coulomb} for $j=5$ (left) and $j=6$ (right).}
\label{potentials}
\end{centering}
\end{figure}

Having established  that both endpoints of  \eqref{sturm} are  limit-points  we can now use SLEIGN2  \cite{Bailey, Bailey2} to compute  eigenvalues. Picking parameter values $L = 1, p = 10, n = -8$ (which satisfy condition \eqref{ESBSC}),  
the potential $Q$  \eqref{ESpotential} occurring in the  differential equation \eqref{sturm} becomes
\begin{equation}
\label{coulomb}
Q(r) =  -\frac{1}{4r^2(r-1)^2} + \frac{(r-5)^2}{(r-1)^2}+\frac{j(j+1)-25}{r(r-1)} .
\end{equation}
We consider the values $j=5,6,7$ (which satisfy \eqref{jCond}), and plot the resulting potential \eqref{coulomb} in Fig.~\ref{potentials} for $j=5$ and $j=6$.  The potential  has a repulsive core and an attractive Coulomb tail. As expected, the minimum becomes shallower and moves outwards as $j$ increases.

Our numerical results for the 11 lowest lying eigenvalues are listed in Table \ref{SLEIGN2} together with the approximation computed via the TN formula  \eqref{lambdaapprox} with $n=-8$, and $\tilde p =10$, i.e.,     
\bee
E(10,4,N) = -8+ 2N\left(\sqrt{N^2+9}-N\right), \quad N  =  5,6,\ldots.
\eee
In the table we use the label $N=5,6,\ldots$ of the TN approximation also to label the numerically computed eigenvalues of \eqref{sturm} with the potential \eqref{coulomb}, so that the  eleven lowest-lying eigenvalues we  compute are labelled by $N=5,\ldots,15$.

\begin{table}[!h]
\centering
\begin{tabular}{|c|c|c|c|c|}
\hline
$N$ &   $E(10,4,N)$ & $ \lambda$ for $j=5$  &   $ \lambda$ for $j=6$ &$ \lambda$ for $j=7$   \\ \hline
 $5$     & 0.3095  &    0.3133  & 0.5107  &  0.6371   \\ \hline
$6$     & 0.4984   & 0.5008  & 0.6290  &   0.7153     \\ \hline
$7$         & 0.6208     & 0.6223   & 0.7097&  0.7711   \\ \hline
$8$      & 0.7041   & 0.7051    &0.7670 &  0.8122   \\ \hline
$9$      & 0.7630  & 0.7637   & 0.8091&  0.8432      \\ \hline
$10$        & 0.8061      & 0.8066   &0.8409&  0.8672    \\ \hline
$11$         & 0.8386   & 0.8390       &0.8654&  0.8861  \\ \hline
$12$          & 0.8636 & 0.8639      &0.8846&   0.9013  \\ \hline
$13$         & 0.8833   & 0.8835    &0.9001&   0.9136   \\ \hline
$14$         & 0.8990    & 0.8992    &0.9127&  0.9238     \\ \hline
$15$     & 0.9118     & 0.9119       &0.9230&  0.9323   \\ \hline
\end{tabular}
\caption{Approximate eigenvalues  computed via  the TN approximation \eqref{Espectrum}  and numerically computed eigenvalues $\lambda$ of the Sturm-Liouville problem \eqref{sturm} 
for $p=10$, $n=-8$  and $j=5,6,7$.}
\label{SLEIGN2}
\end{table}

As expected, the eigenvalues  accumulate at $\lambda = 1$. The TN approximation is remarkably accurate for the smallest allowed value $j=5$, and the accuracy grows   with $N$, reaching $0.01\%$ for the eleventh eigenvalue. The approximation gets significantly worse as $j$ is increased. In all cases, the TN formula give  a smaller value than the numerically computed eigenvalue. Since increasing $j$ makes the potential  shown in Fig.~\ref{potentials} shallower, it is not surprising that eigenvalues increase with $j$. Since the TN approximation is independent  of $j$ it is also not surprising that it is not  equally good for all values of $j$.

\begin{table}[!h]
\centering
\begin{tabular}{|c|c|c|}
\hline
$N$ & $E(6,2,N)$ & $ \lambda$      \\ \hline
 $3$     &0.4499  &    0.4618      \\ \hline
$4$     &  0.6606  &  0.6664     \\ \hline
$5$         &  0.7723   & 0.7754      \\ \hline
$6$      &   0.8375&  0.8394   \\ \hline
$7$      &  0.8786 &    0.8798   \\ \hline
$8$        &  0.9060&  0.9068   \\ \hline
$9$         & 0.9251   & 0.9257  \\ \hline
$10$          &  0.9390 &  0.9395   \\ \hline
$11$         &  0.9494   & 0.9496 \\ \hline
$12$         &  0.9573 &  0.9576      \\ \hline
$13$     & 0.9636 &   0.9637    \\ \hline
\end{tabular}
\qquad \qquad 
\qquad  
\begin{tabular}{|c|c|c|}
\hline
$N$ &   $E(16,7,N)$ & $ \lambda$      \\ \hline
 $8$     &  0.2111 &   0.2123      \\ \hline
$9$     & 0.3633   &    0.3641   \\ \hline
$10$    & 0.4761        &     0.4768  \\ \hline
$11$      &   0.5619 &    0.5624 \\ \hline
$12$      &  0.6285 &    0.6289   \\ \hline
$13$        &  0.6812& 0.6815    \\ \hline
$14$         &  0.7235  &  0.7237 \\ \hline
$15$          &  0.7580 &   0.7582  \\ \hline
$16$         &    0.7865 &   0.7866 \\  \hline
$17$         & 0.8103  &    0.8104    \\ \hline
$18$     & 0.8303 &    0.8304   \\ \hline
\end{tabular}
\caption{Approximate eigenvalues  computed via  the TN approximation \eqref{Espectrum}  and numerically computed eigenvalues $\lambda$ of the Sturm-Liouville problem \eqref{sturm}  for $p=6$, $n=-4$, $j=3$ (left) and $p=16$, $n=-14$, $j=8$ (right).}
\label{SLEIGN3}
\end{table}

We have explored if the TN approximation is generally very good for the smallest allowed value of $j$ by looking at other values of $p$ and $n$. The results are shown in Table \ref{SLEIGN3}, and confirm the same high level of accuracy found in Table \ref{SLEIGN2}.

\section{Conclusion}

The spectral properties of the ES space studied in this paper show remarkable similarities to those of the TN space,  as discussed  in \cite{JS} and \cite{JS2}. On both the  ES and TN spaces, the Dirac operator has no zero-modes and the (scalar) Laplace operator has no bound states. Both manifolds have a 1-parameter family of  (anti-) self-dual and square-integrable two-forms.  Interpreting them as  abelian instantons,  minimal coupling  of the gauge field has the same qualitative effect on the Dirac and scalar Laplace operators in both cases. The Dirac operator acquires a non-trivial kernel,  and the Laplace operator acquires bound states. 

The kernel of the gauged Dirac operator decomposes into irreducible representations of $SU(2)$ in both cases, but here the details differ. On TN, the kernel is the direct sum of all irreducible representations of all  dimensions up to  the greatest integer strictly  smaller
than a suitably defined magnetic flux \cite{JS}. On the ES space, the kernel is the direct sum of $|p|$ copies of the $|p|$-dimensional irreducible representation for  the magnetic  flux $p$ defined in this paper.  The zero-modes can be computed explicitly and take a rather simple form in terms of holomorphic coordinates on the cigar-shaped submanifold which occurs in both spaces. 

The gauged Laplace operator on TN has rather special properties, related to a dynamical  symmetry and resulting in additional degeneracies in the bound states similar to that found in the hydrogen atom \cite{JS2}.  The gauged Laplace operator on the  ES space does not appear to have any dynamical symmetries, and the bound states had to  be computed numerically. Nonetheless, their existence can be understood, as in TN case, in terms of the geometry of the cigar-shaped submanifolds and the form of the magnetic fields  that they inherit from the abelian instantons. Moreover, the eigenvalues can be approximated with surprising accuracy by matching the ES and TN eigenvalue problems, and using the exact formula for TN eigenvalues.  

The accuracy of the TN approximation to the bound state energies of the gauged ES Laplacian  is striking,  and echoes  a similar surprise in the study of bound states of the  Laplace operator on the Atiyah-Hitchin manifold. As shown in \cite{Manton} and  discussed further in  \cite{BSWS}, the Atiyah-Hitchin Laplacian supports bound states even without twisting by a gauge field, with  bound state energies 
 which can be approximated with unreasonable accuracy by the energies of bound states in the negative mass (but ungauged)  TN   Laplacian.

We end with a speculative remark, and a question. The speculation relates to the interpretation of our results in the context of geometric models of matter, where the ES space is a potential model for a neutron,  with the zero-modes  of the Dirac operator modelling the spin degrees of freedom. From this point of view, the occurrence of one spin 1/2  doublet  for $|p|=2$ is welcome, but the presence of two such doublets which are distinguished by their charge under the $U(1)$ symmetry is a puzzle. (Almost) degenerate spin doublets do, of course, occur in nuclear physics in the form of the neutron and the proton. There the degeneracy is usually interpreted in terms of isospin symmetry; here, the origin of the degeneracy is topological (an index). Nonetheless one could speculate that the `doublet of  spin 1/2 doublets' that we have inadvertently found could serve as  model for the quantum states of the  neutron and proton. However, in that case one would have to interpret the $U(1)$ charge which distinguishes the doublets in terms of the charge which distinguishes the proton from the neutron, namely the electric charge. This would be dual to the philosophy of \cite{AMS} where the $U(1)$ symmetry is interpreted as a magnetic symmetry, dual to the $U(1)$ of electromagnetism. 

The question  that we would like to raise is what our results mean in the context of (quantum) gravity. The Lorentzian Schwarzschild solution is the most studied solution of the Einstein equations, and the various wave equations on it have equally been under intense scrutiny. The Euclidean Schwarzschild solution is  the simplest `gravitational instanton' and, as such, plays a key role in semiclassical approaches to quantum gravity. The periodicity of its `Euclidean time coordinate' can be related to the thermal nature of black body radiation \cite{Wald}. What, then,  is the interpretation of the abelian instanton on the ES space,  and of the bound states which the Dirac and Laplace operators develop when twisted by the instanton?

\vspace{0.2cm}

\noindent {\bf Acknowledgments} \;\; RJ thanks MACS at  Heriot-Watt University for a PhD scholarship. BJS acknowledges support through the EPSRC grant `Dynamics in Geometric Models of Matter' (EP/K00848X/1),  and thanks Gary Gibbons for correspondence and for pointing out reference \cite{Fawcett}. This research was supported in part by the Perimeter Institute for Theoretical Physics. Research at the Perimeter Institute is supported by the Government of Canada through Industry Canada and by the Province of Ontario through the Ministry of Economic Development and Innovation.

\vspace{1cm}
\appendix
\noindent {\bf\large  Appendix}

\section{The gauged Laplace operator on the two-sphere in complex coordinates}
\label{s2app}

In terms of the  stereographic complex coordinates \eqref{EScomplex} on the  two-sphere without the south pole, the metric on the  two-sphere of unit radius is 
\bee
ds^2 = \frac{4}{q^{2}} dz d\bar z, \qquad q=1+z\bar z,
\eee
and the gauge potential of the Dirac monopole of charge $p$ is 
\bee
\label{Diracp}
A^p_N=\frac{p}{2q}(zd\bar z -\bar z dz).
\eee
The Laplace operator on the two-sphere in terms of these coordinates is simply
\bee
\Delta_S^2 = q^2\partial_z\bar \partial_z = \frac{q^2}{2} (\partial_z\bar \partial_z 
+  \bar \partial_z \partial_z).
\eee
Minimal coupling to \eqref{Diracp} gives
\begin{align}
\label{s2lap}
\Delta_{S^2,p} &=\frac{q^2}{2}\left( \left(\partial_z-\frac{p}{2q}\bar z\right) \left(\bar \partial_z + \frac{p}{2q}z\right) + \left(\bar \partial_z + \frac{p}{2q}z\right)
 \left(\partial_z-\frac{p}{2q}\bar z\right)\right) \nonumber \\
 &= \frac{1}{2}\left( \left(q\partial_z-\frac{p+2}{2}\bar z\right) \left(q\bar \partial_z + \frac{p}{2}z\right) + \left(q\bar \partial_z + \frac{p-2}{2}z\right)
 \left(q\partial_z-\frac{p}{2}\bar z\right)\right).
 \end{align}
 Using  the relations \eqref{ESdd} we can also write 
 \bee
\Delta_{S^2,p}  = \frac{1}{2}\left( q^{\frac{p}{2}+2}\partial_z q^{-p}\bar \partial_z q^{\frac{p}{2}} +  
  q^{-\frac{p}{2}+2}\bar \partial_z q^p \partial_z q^{-\frac{p}{2}} \right).
\eee

Adopting the conventions of \cite{JS}, we  write $H$ for the hyperplane bundle over $S^2\simeq \mathbb{CP}^1$, which is the dual bundle to the tautological bundler over $ \mathbb{CP}^1$.  Then, denoting the space of  sections of  the $p$-th power  $H^p$ by $C^\infty(H^p)$, we have 
\bee
q\partial_z-\frac p 2 \bar z: C^\infty(H^p)\rightarrow C^\infty(H^{p-2}), \qquad 
 q\bar \partial_z + \frac{p}{2}z: C^\infty(H^p)\rightarrow C^\infty(H^{p+2}), 
\eee
so that 
\bee
\Delta_{S^2,p} : C^\infty(H^p)\rightarrow C^\infty(H^{p}).
\eee

As discussed in detail in \cite{JS}, we can obtain local sections of $H^p$ by pulling back ordinary functions on $S^3$ obeying a constraint. Working with such equivariant functions has several advantages: there is no need to introduce local sections, and many of the interesting differential operators acting on sections of $H^p$ are naturally associated to $S^3\simeq SU(2)$. Referring the reader to \cite{JS} for details,   we use  complex coordinates $(z_1,z_2)$ satisfying the constraint $|z_1|^2+|z_2|^2 =1$ to parametrise $S^3$. 

In terms of these coordinates, the $SU(2)$ right-action on itself has the generators
\begin{align}
X_+& =i (z_1\bar{\partial}_2 -z_2 \bar{\partial}_1), \nonumber  \\
X_-& =i (\bar{z}_2\partial_1 -\bar{z}_1 \partial_2), \nonumber \\
X_3& =\frac  i  2
 ( \bar{z}_1\bar{\partial}_1 +\bar{z}_2 \bar{\partial}_2 - z_1 \partial_1 -z_2 \partial_2),
\end{align}
while the left-action is generated by
\begin{align}
\label{leftgenerators}
Z_+& = i(z_2\partial_1 -\bar{z}_1\bar{\partial}_2), \nonumber \\
Z_- & = i(z_1\partial_2 -\bar{z}_2\bar{\partial}_1), \nonumber \\
Z_3& =\frac i 2  (z_1 \partial_1 - z_2 \partial_2 - \bar{z}_1\bar{\partial}_1 +\bar{z}_2 \bar{\partial}_2   ).
\end{align}

The complex coordinate $z$ parametrising $S^2$ without the south pole is related to $(z_1,z_2)$ via $z=z_2/z_1$, and we can use the following local section of the Hopf bundle
\bee
\label{localsect}
s_N:\CC \rightarrow S^3 \quad z\mapsto \frac{1}{\sqrt{q}}(1,z)
\eee
for pulling back functions on $S^3$ to local sections of line bundles over $S^2$.
Defining the space of equivariant functions
\bee
C^\infty(S^3,\CC)_s= \{F: S^3 \rightarrow \CC|iX_3F=sF\},
\eee
the results of \cite{JS}  then imply  the following commutative diagrams involving the pull-back  via \eqref{localsect}:
\bee
\begin{CD}
C^\infty(S^3,\CC)_\frac{p}{2}  @>X_+>> C^\infty(S^3,\CC)_{\frac{p+2}{2}} \\
@VV s*_N V   @VV s*_N V \\
 C^\infty(H^p) @> i(q\bar \partial_z + \frac p 2 z)>>  C^\infty(H^{p+2}),
\end{CD}
\eee
and 
\bee
\begin{CD}
C^\infty(S^3,\CC)_\frac{p}{2}  @>X_->> C^\infty(S^3,\CC)_{\frac{p-2}{2}} \\
@VV s*_N V   @VV s*_N V \\
 C^\infty(H^p) @> -i(q \partial_z -  \frac p 2 \bar z)>>  C^\infty(H^{p-2}).
\end{CD}
\eee

Since the Laplacian on the 3-sphere can be written as 
\bee
\label{S2S3}
\Delta_{S^3}= \frac 12 \left( X_-X_+  + X_+ X_- \right) +  X_3^2,
\eee
the expression \eqref{s2lap}  for  $\Delta_{S^2,p} $  gives rise to the  commutative diagram
\bee
\label{comdiagram}
\begin{CD}
C^\infty(S^3,\CC)_\frac{p}{2}  @>\frac 12 \left( X_-X_+  + X_+ X_- \right)  >> C^\infty(S^3,\CC)_{\frac{p}{2}} \\
@VV s*_N V   @VV s*_N V \\
 C^\infty(H^p) @> \Delta_{S^2,p} >>  C^\infty(H^{p}).
\end{CD}
\eee
This shows that we can obtain eigensections of $\Delta_{S^2,p} $ by pulling back simultaneous eigenfunctions of $ \Delta_{S^3}$ (total angular momentum)  and $X_3^2$ (body-fixed angular momentum).

To obtain a unique labelling of eigenfunctions in terms of eigenvalues we include the operator $Z_3$ (space-fixed angular momentum). One checks that, for
$F\in C^\infty(S^3,\CC)_\frac{p}{2}$, 
\bee
iZ_3F=\left(-\frac{p}{2}  + z_2\partial_2-\bar{z}_2\bar{\partial}_2\right)F. 
\eee
Using this, we can relate $Z_3$ to the  generator of rotations about the `vertical' axis of $S^2$
\bee
L_3 = \partial_\phi =i(z\partial_z -\bar{z}\bar{\partial}_z),
\eee
where we used the angular coordinate defined in \eqref{EScomplex}. The relation is again best shown in a commutative diagram:
\bee
\label{comL3diagram}
\begin{CD}
C^\infty(S^3,\CC)_\frac{p}{2}  @>-iZ_3 -\frac{p}{2}  >> C^\infty(S^3,\CC)_{\frac{p}{2}} \\
@VV s*_N V   @VV s*_N V \\
 C^\infty(H^p) @> iL_3 >>  C^\infty(H^{p}).
\end{CD}
\eee

As explained in \cite{JS} and \cite{JS2}, one  obtains simultaneous eigenfunctions of the  Hermitian operators $\Delta_{S^3},$  $ iZ_3$ and  $iX_3$ from polynomials in $z_1,z_2,\bar{z}_1,\bar{z}_2$ which are annihilated by the four-dimensional Laplacian $\Box_4= 4(\partial_1\bar{\partial}_1 + \partial_2\bar{\partial}_2)$. One checks easily that the following polynomials are in the kernel of $\Box_4$:
\begin{equation}
\label{Y}
Y^j_{sm} =C_{jms}
\sum_k\frac{  (-1)^{-k} }{(j+m-k)!k!(j-s -k)!(s -m + k)!}z_1^{s-m+k}z_2^{j+m-k}\bar{z}_1^k\bar{z}_2^{j-s-k},
\end{equation}
where  $C_{jms}$ is an overall normalisation constant
\bee
C_{jms}=(-1)^{j-s}  \left((j+s)!(j-s)!(j+m)!(j-m)!\right)^{\frac 12 },
\eee
and
\bee
j\in\frac 1 2 \NN^0, \quad s, m=-j,-j+1,\ldots,j-1,j.
\eee
The summation index  $k$ runs over the values so that the factorials are well defined.
These functions are orthonormal and satisfy
\begin{equation}
\label{FI}
\Delta_{S^3}Y^j_{sm} = -j(j+1)Y^j_{sm}, \ \ iZ_3Y^j_{sm} = mY^j_{sm}, \ \ iX_3Y^j_{sm} = sY^j_{sm}.
\end{equation}

It now follows from the relation \eqref{S2S3} and the commutative diagram \eqref{comdiagram} that
 the pullbacks of the functions \eqref{Y} to the two-sphere without the south pole
 \bee
 \label{mopoharmonics}
 y^{j}_{\frac p 2 m} = s^*_N Y^j_{\frac p 2 m} 
 \eee
satisfy  
\bee
\Delta_{S^2,p}  y^{j}_{\frac p 2 m}= \left(-j(j+1) +\frac{p^2}{4}\right) y^{j}_{\frac p 2 m},
\eee
and  
\bee
iL_3  y^{j}_{\frac p 2 m}=-\left(m +\frac p2\right)y^{j}_{\frac p 2 m}.
\eee

The special cases $s=\pm j$ occur as the angular dependence of  the zero-modes of the Dirac operator discussed in the main text. The summations in \eqref{Y} collapse to a single term.  The case $s=j$  only allows for $k=0$,  giving the holomorphic functions
\bee
Y^j_{jm} =\sqrt{\frac{ (2j)!}{(j+m)!(j -m)!}}z_1^{j-m}z_2^{j+m},
\eee
while in  the case $s=-j$ the  only allowed term is $k=j+m$, producing the anti-holomorphic functions
\bee
Y^j_{(-j)m} =(-1)^{j-m}\sqrt{\frac{ (2j)!}{(j+m)!(j -m)!}}\bar{z}_1^{j+m}\bar{z}_2^{j-m}.
\eee
The pull-backs for $s=j$, are now 
\bee
\label{yjm}
 y^{j}_{j m} = \sqrt{\frac{ (2j)!}{(j+m)!(j -m)!}}q^{-j }z^{j+m}, 
\eee
while for $s=-j$ we have 
\bee
\label{y-jm}
 y^{j}_{(-j) m} = (-1)^{j-m}\sqrt{\frac{ (2j)!}{(j+m)!(j -m)!}}q^{-j }\bar{z}^{j-m}.
\eee
With  $m$ ranging over $-j,-j+1,\ldots j-1, j$, both of these span $(2j+1)$-dimensional irreducible representations of $SU(2)$, see \cite{JS} for details.


\begin{thebibliography}{99}

\bibitem{Hawking} 
S.~W.~Hawking, Gravitational instantons, Phys.~Lett.~A 60 (1977) 81--83.

\bibitem{GH} G.~W.~Gibbons and S.~W.~Hawking, Classification of gravitational instanton symmetries, Commun.~Math.~Phys. 66 (1979)
291--310.

 \bibitem{AMS} M.~Atiyah, N.~S.~Manton and B.~J.~Schroers, Geometric models of matter,  Proc.~Roy.~Soc.  A468 (2012)  1252--1279. 


 \bibitem{AFS} M.~Atiyah, N.~Franchetti and B.~J.~Schroers, Time evolution in a geometric model of a particle,  JHEP 02 (2015) 062.
 

 \bibitem{GM} G.~W. ~Gibbons and N.~S. ~Manton, Classical and quantum dynamics of BPS monopoles, Nucl.~Phys. B274  (1986) 183--264. 
 
\bibitem{Manton} N.~S.~Manton, Monopole and Skyrmion bound states, Phys.~Lett.~B 198 (1987) 226--230.
  
\bibitem{Schroers} B.~J.~Schroers, Quantum scattering of BPS monopoles at low energy, Nucl.~Phys.~B367 (1991) 177--214 . 
  
\bibitem{JS} R.~Jante and B.~J.~Schroers, Dirac operators on the Taub-NUT space, monopoles and $SU(2)$ representations,
  JHEP, { 1401} ( 2014) 114.
  
   
\bibitem{JS2} R.~Jante and B.~J.~Schroers, Taub-NUT dynamics with a magnetic field, J.~Geom.~Phys. 104 (2016) 305--328.

\bibitem{Pope1} C.~N.~Pope, Axial-vector anomalies and the index theorem in charged Schwarzschild and Taub-NUT spaces, Nucl.~Phys.~B141 (1978), 432--444.


\bibitem{Etesi}G.~Etesi and T.~Hausel, Geometric interpretation of Schwarzschild instantons, J.~Geom.~Phys.   37 (2001) 126--136.

\bibitem{Fawcett} M.~S.~Fawcett,  The energy-momentum tensor near a black hole,  Commun.~Math.~Phys. 89 (1983) 103--115.

\bibitem{Wald} R.~M.~Wald, General relativity, The University of Chicago Press, 1984.


\bibitem{NP} E.~T. Newman and R.~Penrose, Note on the Bondi-Metzner-Sachs group, 
J.~Math.~Phys. 7 (1966)  863--879.

\bibitem{Dray2}   T.~Dray, A unified treatment of Wigner $\mathcal{D}$ functions, spin-weighted spherical harmonics, and monopole harmonics, J.~Math.~Phys. 27 (1986) 781--792.
  

     


\bibitem{Jost} J.~Jost, Riemannian Geometry and geometric analysis, 2nd ed., Springer, Berlin, 1998.
  
\bibitem{Bailey} P.~B.~Bailey, W.~N.~Everitt and A.~Zettl, Computing eigenvalues of singular Sturm-Liouville problems, Results in Math. 20 (1991)  391--423. 

 \bibitem{Bailey2} P.~B.~Bailey, W.~N.~Everitt and A.~Zettl, The SLEIGN2 Sturm-Liouville code. ACM Trans. Math. Software 27 (2001) 143--192. 
  
\bibitem{RS} M.~Reed and S.~Simon, Methods of modern mathematical physics, Vol. II: Fourier Analysis, Self-Adjointness, Academic Press, 1975; Vol. IV: Analysis of Operators,  Academic Press, 1978.
 
\bibitem{BSWS} L.~Boulton, B.~J.~Schroers  and K.~ Smedley-Williams, Quantum bound states in Yang-Mills-Higgs theory at critical coupling, in preparation.




  
 


\end{thebibliography}
\end{document}